\documentclass[floatfix,reprint,amsmath,amssymb,aps,prb,twocolumn,superscriptaddress]{revtex4-2}

\usepackage{graphicx,units}
\usepackage{dcolumn}
\usepackage{bm}
\usepackage{bbold}
\usepackage[plainpages=false,pdfpagelabels,colorlinks=true,linkcolor=red,urlcolor=blue,citecolor=blue,pdftitle={Title},pdfauthor={},pdfdisplaydoctitle=true,pdfduplex=DuplexFlipLongEdge]{hyperref}
\usepackage[caption=false]{subfig}
\usepackage{xcolor}
\usepackage{physics}
\usepackage{resizegather}
\usepackage[T1]{fontenc} 
\usepackage[normalem]{ulem} 

\graphicspath{{./figures/}} 

\newcommand{\pcsadd}{Centre for Theoretical Physics of Complex Systems, Institute for Basic Science (IBS), Daejeon 34126, Republic of Korea}
\newcommand{\ustadd}{Basic Science Program, Korea University of Science and Technology (UST), Daejeon, Korea, 34113}

\newcommand{\vect}[1]{\mathbf{#1}}

\newcommand{\mh}{\hat{\mathcal{H}}}
\newcommand{\mU}{\hat{\mathcal{U}}}
\newcommand{\mD}{\hat{\mathcal{D}}}

\newcommand{\mhon}{\mh_\mathrm{onsite}}
\newcommand{\mdon}{\mD_\mathrm{onsite}}
\newcommand{\mhrmf}{\mh_\mathrm{RMF}}
\newcommand{\mhsoc}{\mh_\mathrm{SOC}}
\newcommand{\mhsoco}{\mh_\mathrm{SOCO}}

\newcommand{\Pa}{\hat{P}_a}
\newcommand{\tphi}{\tilde{\phi}}

\newcommand{\hd}{\hat{d}}
\newcommand{\hp}{\hat{p}}
\newcommand{\hf}{\hat{f}}

\newcommand{\apn}{\langle\mathrm{PN}\rangle}
\newcommand{\ttau}{\tilde{\tau}}
\newcommand{\tE}{\tilde{E}}
\newcommand{\bE}{\bar{E}}

\newcommand{\om}{\omega}
\newcommand{\Th}{\theta}

\begin{document}

\title{Flat Band Induced Metal-Insulator Transitions for Weak Magnetic Flux and Spin-Orbit Disorder}

\author{Yeongjun Kim}
    \email{yeongjun.kim.04@gmail.com}
    \affiliation{\pcsadd}
    \affiliation{\ustadd}

\author{Tilen \v{C}ade\v{z} }
    \email{tilencadez@ibs.re.kr}
    \affiliation{\pcsadd}

\author{Alexei Andreanov}
    \email{aalexei@ibs.re.kr}
    \affiliation{\pcsadd}
    \affiliation{\ustadd}

\author{Sergej Flach}
    \email{sflach@ibs.re.kr}
    \affiliation{\pcsadd}
    \affiliation{\ustadd}

\date{\today}

\begin{abstract}
    We consider manifolds of tunable all-bands-flat (ABF) lattices in dimensions \(d=1,2\), parameterized by a manifold angle parameter \(\theta\).
    We study localization properties of eigenstates in the presence of weak magnetic flux disorder and weak spin-orbit disorder.
    We demonstrate that weakly disordered ABF lattices are described by effective scale-free models, where the disorder strength is scaled out.
    For weak magnetic flux disorder, we observe sub-exponential localization at flatband energies in \(d=1\), which differs from the usual Anderson localization.
    We also find diverging localization length at flatband energies for weak flux values in \(d=2\), however the character of the eigenstates at these energies is less clear.
    For weak spin-orbit coupling disorder in \(d=2\) we identify a tunable metal-insulator transition with mobility edges.
    We also consider the case of mixed spin-orbit and diagonal disorder and obtain the metal-insulator transition driven by the manifold parameter \(\theta\).
\end{abstract}

\maketitle

\section{Introduction}
\label{sec:intro}

There is an increasing interest in flatband systems (flatbands)~\cite{parameswaran2013fractional, bergholtz2013topological, derzhko2015strongly, leykam2018artificial, leykam2018perspective} in the fields of photonics and condensed matter physics.
Flatbands arise in tight-binding lattices where one or more energy bands \(E(\vect{k})\) are dispersionless.
They were initially found in certain lattice systems as fine-tuned cases~\cite{weaire1971eletronic} or due to bipartite symmetry~\cite{sutherland1986localization, lieb1989two}. 
Soon thereafter, more systematic approaches to construct them were developed, based on line graphs~\cite{mielke1991ferromagnetic, mielke1991ferromagnetism} and decorated lattices~\cite{tasaki1992ferromagnetism}. 
Further the impact of chiral symmetry on the existence of flatbands was explored~\cite{shima1993electronic, lan2012coexistence, zhong2017transport, ramachandran2017chiral} and more recently also other types of symmetries leading to protected flatbands, 
such as latent~\cite{roentgen2018compact,morfonios2021latent} and anti-\(\mathcal{PT}\)~\cite{mallick2022antipt}, were proposed. 
New systematic approaches to flatband construction based on either real~\cite{maimaiti2017compact, toikka2018necessary, maimaiti2019universal, maimaiti2021flatband} or momentum space~\cite{rhim2019classification, hwang2021general, graf2021designing} were introduced.
These works demonstrate that flatband models form continuous manifolds and their properties can be tuned by changing the manifold parameters.
In this work we focus on an interesting extreme case of flatband lattices where all the bands are flat (all-bands-flat, ABF)~\cite{vidal1998aharonov, danieli2021nonlinear}. 
The absence of dispersive bands in an ABF model makes all the eigenstates localized and the system completely insulating. 

Due to the macroscopic degeneracy, perturbed flatband lattices exhibit interesting phenomena, such as ferromagnetism~\cite{derzhko2015strongly, lieb1989two, mielke1991ferromagnetic, mielke1991ferromagnetism, tasaki1992ferromagnetism, mielke1993ferromagnetism, ramirez1994strongly, mielke1999stability}, 
superfluidity and superconductivity~\cite{peotta2015superfluidity, julku2016geometric, tovmasyan2018preformed, mondaini2018pairing, aoki2020theoretical}, many-body localization~\cite{kuno2020flat, danieli2020many, vakulchyk2021heat, danieli2022many}, 
and unconventional Anderson localization~\cite{vidal2001disorder, goda2006inverse, nishino2007flat, chalker2010anderson, leykam2013flat, flach2014detangling, leykam2017localization, ramachandran2017chiral, shukla2018disorder, longhi2021inverse}. 
The perturbations can be various types of disorder or interactions, and in this work we are interested in the former. 
As for the latter, the non-perturbative effect of weak interaction on many-body ABF can lead to a nontrivial transport of pairs, topological effects and quantum caging~\cite{tovmasyan2013geometry, takayoshi2013phase, pelegri2020interaction, kuno2020interaction, nicolau2023many}.

In lattices with only dispersive bands, Anderson showed that strong onsite disorder leads to complete localization of the eigenstates~\cite{anderson1958absence}. 
For weak disorder, the system can be treated perturbatively in momentum space, making it diffusive in \(d = 3\). 
Because the kinetic energy is quenched in a flatband, one may conclude that flatband systems are trivially localized, since the disorder is effectively infinite.
However, the situation is not as simple as one might expect. 
Infinitesimal disorder breaks the fine-tuned flatness and destructive interference condition for the compact localized states (CLS), and can delocalize the eigenstates. 
Indeed, in our recent work~\cite{cadez2021metalinsulator} we have derived effective Hamiltonians of ABF lattices perturbed with infinitesimally weak onsite disorder and then projected on a flatband Hilbert space.
We have shown that the eigenstates of the effective model evolve non-perturbatively, inducing a metal-insulator transition(MIT) in \(d = 3\), while the system remains localized in \(d = 1, 2\).

The Anderson transition is a quantum phase transition, showing universal behavior~\cite{kramer1993localization, evers2008anderson}.
In its vicinity, there is a diverging localization length \(\xi\) with a critical exponent \(\nu\).
The localization properties are determined by the symmetries and the dimensionality of the system.
An important insight into the details of the Anderson transition is based on renormalization arguments and a one parameter scaling hypothesis~\cite{abrahams1979scaling}.
The scaling theory of localization predicts MIT in \(d = 3\) and only localized states in \(d = 1\). 

The \(d=2\) case is marginal~\cite{hikami1980spin}, and the existence of MIT in \(d=2\) has been studied extensively.
In Ref.~\onlinecite{hikami1980spin}, the RG analysis of the nonlinear \(\sigma\)-model established that out of the three Wigner-Dyson symmetry classes: orthogonal, unitary, symplectic classes, the MIT in \(d = 2\) can occur only for the symplectic class.
This prediction has been verified numerically~\cite{asada2002anderson, asada2004numerical}. 
However, it was found that MIT in \(d=2\) occurs in the presence of the strong constant magnetic field, also known as the quantum Hall transition~\cite{slevin2009critical}.
The issue extends beyond the standard Wigner-Dyson symmetry classes~\cite{altland1997nonstandard}, where unconventional localization can occur in \(d = 1, 2\)~\cite{evers2008anderson}.
Two notable and most studied examples are the chiral orthogonal and unitary classes.
These systems have bipartite lattices with disordered hoppings. 
They exhibit a symmetric energy spectrum in the presence of disorder, with an \(E = 0\) eigenstate for odd number of sites. 
An analytical study~\cite{fabrizio2000anderson} predicts that these systems exhibit a localization length which diverges as the logarithm of \(E\) in the vicinity of \(E = 0\), with a completely delocalized \(E = 0\) eigenstate.
For \(d = 1\), studies on chains with hopping disorder have shown a diverging localization length near \(E = 0\), while the eigenstate precisely at the band center is sub-exponentially localized~\cite{fleishman1977fluctuations, inui1994unusual}.
In \(d = 2\) numerical studies of models with hopping disorder and random magnetic flux (RMF) on a square lattice also found a divergence of the localization length around \(E = 0\).
Several works \cite{ziman1982localziation, inui1994unusual, xiong2001power} have studied the localization of the \(E = 0\) eigenstate for \(d = 2\) and arrived at conflicting conclusions on the localization properties of the state.
The divergent density of states around \(E=0\) makes numerical computations and characterizations of the state challenging.
However we point out that recently an MIT was identified in a 2D system subject to RMF~\cite{li2022transition}.

In this work, we extend our previous results for onsite disorder~\cite{cadez2021metalinsulator} and study the effects of different types of infinitesimally weak disorder on ABF, namely the random magnetic flux (RMF) and the random spin-orbit couplings (SOC).
This is motivated by the peculiar and interesting properties of dispersive 2D systems with these types of disorder, as mentioned above.
We show that when ABF is perturbed by an infinitesimal RMF disorder, the effective non-perturbative model exhibits a particle-hole symmetry and displays spectral properties similar to those of bipartite lattices mentioned in the preceding paragraph.
We show analytically and numerically that an infinitesimal RMF leads to anomalous localization in \(d = 1, 2\), with the existence of a sub-exponentially localized state at zero energy.
In the case of the SOC disorder, we report on a tunable non-perturbative MIT with mobility edges in \(d = 2\).
The paper is organized as follows: we introduce the theoretical background of the scale-free model in Sec. \ref{sec:model}.
We study RMF and SOC disorder in Sec.~\ref{sec:rmf} and~\ref{sec:soc}, respectively.
Each section contains numerical results. This is followed by conclusions in Sec. \ref{sec:conclusion}.

\section{Construction of ABF networks and scale-free models}
\label{sec:model}

We begin by outlining the ABF construction procedure~\cite{danieli2021nonlinear, cadez2021metalinsulator} for \(d = 1, 2\) ABF lattices with \(\nu = 2\) flatbands and nearest unit cell hoppings~\cite{maimaiti2021flatband}.
The starting point is a lattice of decoupled sites.
The sites on each sublattice \(a,b\) are assigned the same onsite energy: \(E_a = -\Delta/2\) and \(E_b = \Delta/2\), so that the bandgap is \(\Delta\).
Therefore, the Hamiltonian has a trivial diagonal representation in real space in this basis, which we refer to as the \emph{fully-detangled basis}.
In this basis, the ABF Hamiltonian \(\mh\) reads
\begin{align}
    \label{eq:fd}
    \mh = \frac{\Delta}{2} \sum_{\vect{r}} -\ketbra{a, \vect{r}}{a, \vect{r}} + \ketbra{b, \vect{r}}{b, \vect{r}}
\end{align}
where \(\ket{a, r}\) and \(\ket{b, r}\) are  detangled (diagonalized) eigenstates in real space. 
From the fully-detangled basis, an ABF Hamiltonian is constructed by the change of basis: applying a finite sequence of \((d+1)\) non-commuting local unitary transformations (LUT).
Reversing the above procedure implies these ABF Hamiltonians can be diagonalized in real space by a finite sequence of local unitary transformations without resorting to the non-local Bloch space representation.
These results hold for all 1D ABF Hamiltonians~\cite{sathe2020compactly,danieli2021nonlinear} and we believe that it is also true in higher dimensions~\cite{danieli2021nonlinear}.

In the simplest case, LUTs couple each unit cell with an element of the \(\mathrm{SO}(2)\) group as follows
\begin{align}
    \label{eq:lut}
    \begin{split}
        \mU^{(i)} = \sum_{\vect{r}} \cos \Th_i \ketbra{a, \vect{r}}{a, \vect{r}} + \sin \Th_i \ketbra{a, \vect{r}}{b, \vect{r}_i} \\
        - \sin \Th_i \ketbra{b, \vect{r}_i}{a, \vect{r}} + \cos \Th_i \ketbra{b, \vect{r}_i}{b, \vect{r}_i},
    \end{split}
\end{align}
where \(\theta_i\) are the ABF manifold angle parameters.
The most generic LUT is an element of the \(\mathrm{SU}(2)\) group, parameterized by three angles. 
The unit cell redefinition is required to couple the nearest unit cells in \(d\) dimensions (for concreteness, Eq.~\eqref{eq:lut} illustrates the \(d=2\) case) given by
\begin{align*}
    \vect{r}_i &= \vect{r} + \delta_{i,1}\vect{x} +  \delta_{i,2}\vect{y},
\end{align*}
where \(\vect{x}, \vect{y}\) are primitive lattice vectors.
With the combined LUTs \(\mU = \prod_{i=0}^{d}\mU^{(i)}\), we define the \emph{fully-entangled} basis
\begin{align}
    \label{eq:fut}
    \ket{p, \vect{r}} = \mU\ket{a, \vect{r}}, \quad \ket{f, \vect{r}} = \mU\ket{b, \vect{r}}.
\end{align}
In this new basis the ABF Hamiltonian \(\mh\)~\eqref{eq:fd} has \(d\)-dimensional nearest neighbor hoppings.
The expression for the \(d = 1, 2\) ABF Hamiltonian in the fully-entangled basis with \(\theta_i = \theta\) is given in Appendix~\ref{app:fe_2d} and illustrated in Fig.~\ref{fig:fig1}.
We point out that since any unitary transformation can be decomposed into a finite sequence of embedded \(2\times 2\) unitary transformations, the above construction readily extends to the case of a large number of bands \(\nu\).

The eigenstates in the fully-detangled basis occupy a single lattice site, and these eigenstates remain strictly local in the fully-entangled basis thanks to the locality of the basis change transformation.
Let us add a weak disorder \(W\mD\) to the ABF Hamiltonian (in the fully-entangled basis), with strength \(W \ll \Delta\), and an operator \(\mD\).
The degeneracy of flatbands is lifted by the disorder, yet the broadening of spectrum is much smaller than the bandgap \(\Delta\). 
For this reason, the perturbed eigenstates emerging from one of the flatbands (with energy \(E_a\) for concreteness) are effectively described by the first order degenerate perturbation theory
\begin{align}
    \label{eq:sf}
    W \Pa \mD \Pa \ket{\psi} = \delta E \ket{\psi}
\end{align}
where the projector on the flatband with energy \(E_a\) is given by \(\hat{P}_a = \sum_n \ketbra{a, \vect{r}}{a, \vect{r}}\) (similarly for \(\hat{P}_b\)).
We observe that the disorder strength \(W\) becomes simply an overall scaling factor and the above eigenproblem defines a new effective Hamiltonian. 
We scale out \(W\) from the both sides of Eq.~\eqref{eq:sf} and denote it as the scale-free model \(\tilde{\mh} = \Pa \mD \Pa\) with eigenenergies \(\tE = \delta E/W\).
We note that even infinitesimal values of \(W\) always alter the system, reflecting the non-perturbative effect of disorder on the macroscopically degenerate flatband eigenstates.
In what follows, we drop the tilde symbols and focus exclusively on the properties of the scale-free models and their spectrum.
Provided the contribution from the other band, \(b\), is eliminated by the projection and we are effectively dealing with a chain or square lattice, it is more convenient to use the integer unit cell coordinates \(\mathbf{n}\), rather than the vectors \(\vect{r}\).

We note that this projection method is applicable also in the presence of interactions: several works described the non-perturbative effect of an interaction via an effective Hamiltonian, e.g. a projection onto the flatband~\cite{tovmasyan2013geometry,takayoshi2013phase,pelegri2020interaction,kuno2020interaction}.

\begin{figure}
	\includegraphics[width=\columnwidth]{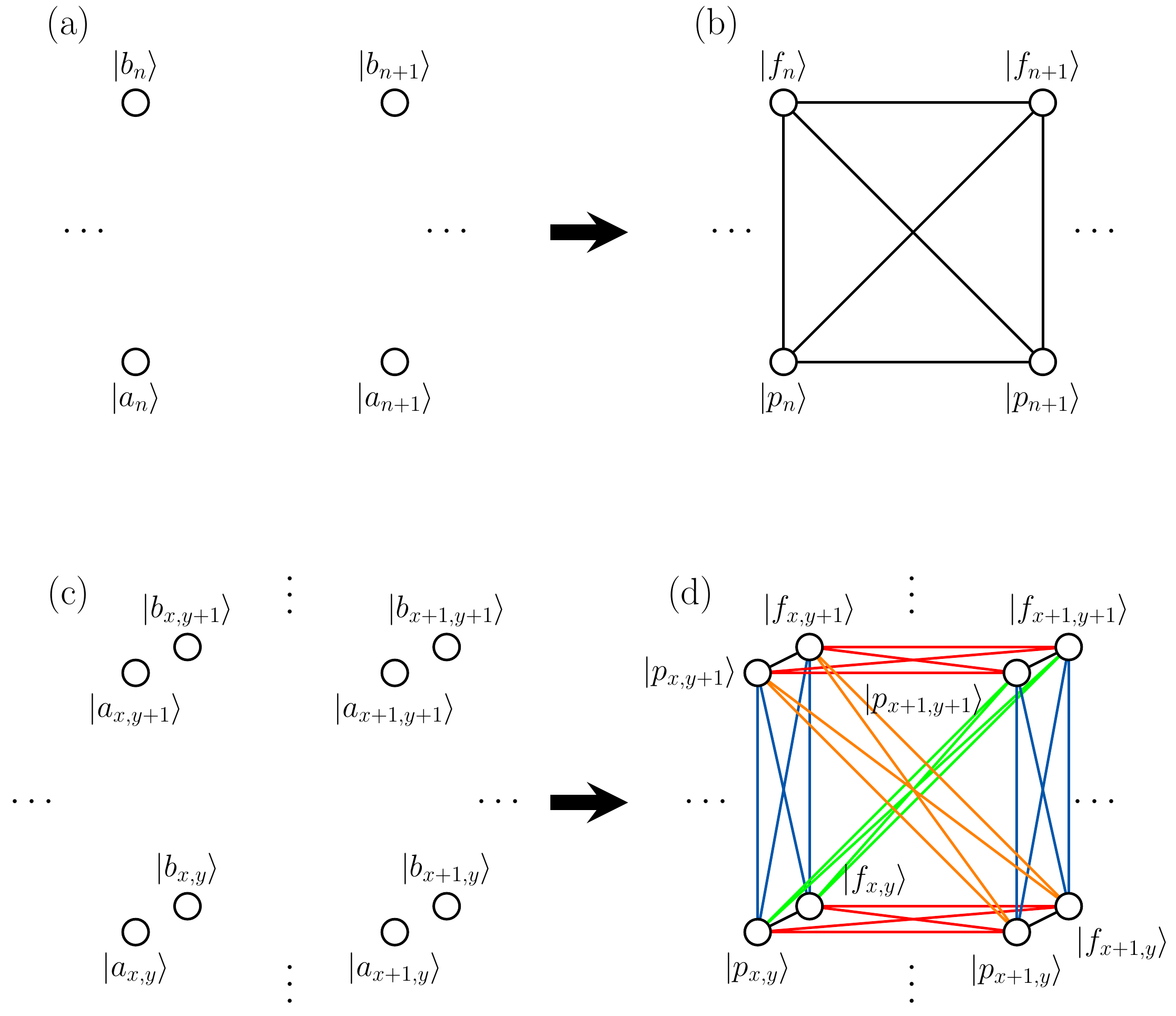} \\
    \caption{
        Schematic representations of the ABF Hamiltonians \(\mh\)~\eqref{eq:fd} in the fully-detangled basis in (a) \(d = 1\) and (c) \(d = 2\) and the fully entangled basis in (b) \(d = 1\) and (d) \(d = 2\). 
        In the 2D case the color of the hoppings represents \(5\) different hopping matrices~\cite{maimaiti2021flatband}: \(H_0\) (black), \(H_{(1, 0)}\) (red), \(H_{(0, 1)}\) (blue), \(H_{(1, 1)}\) (green), \(H_{(1, -1)}\) (orange), which are given explicitly in App.~\ref{app:fe_2d}.
        The dots at the edges indicate that the pattern repeats in those directions.
    }
	\label{fig:fig1}
\end{figure}

\subsection{Numerics}
\label{sec:num}

In the next sections we consider two different types of disorder: random magnetic flux (RMF), and random spin-orbit coupling (SOC), the latter both in the presence and the absence of an onsite potential disorder.
The case of onsite disorder was studied in detail before~\cite{cadez2021metalinsulator} and in this work we consider it together with the random SOC.

We construct \(d=1\) and \(d=2\) fully-entangled ABF Hamiltonians with \(\nu = 2\) bands following the procedure outlined above, with a band gap \(\Delta = 2\) as a function of angle \(\theta = \theta_i\) and of system sizes \(2L\)  and \(2L^2\), respectively.
The resulting models are given in App.~\ref{app:fe_2d}.
The \(\Th\) controls the coupling strength, or hopping strength.
Maximum hopping, and the CLS which are maximally expanded in the fully-detangled basis are reached at \(\Th = \pi/4\), while as expected, there is no coupling at \(\Th = 0\).
From \(\mh(\theta)\) we derive the scale-free models \(\mhrmf, \mhsoc\) for infinitesimal RMF and SOC disorder using Eq.~\eqref{eq:sf}.
For \(d = 1\) we present the explicit Hamiltonian for the scale-free model.
The scale-free Hamiltonian for \(d=2\) is lengthy, and we do not derive it explicitly, but rather the unitary transformations and projection were performed numerically.

We study the spectra of these scale-free Hamiltonians using the full diagonalization and the Lanczos method with shift-and-invert technique to obtain eigenstates in a window \(\Delta E\) around the target energy \(E\). 
The Lanczos method enables us to go to larger \(L\) as compared to the full diagonalization, which proves especially useful in \(d = 2\).
The width of the window \(\Delta E\) is chosen so that the quantities averaged over this energy window do not change significantly by choosing a smaller window size.
The appropriate width of the energy window \(\Delta E\) has an almost constant density of states (DoS) over that window.
As we see in Sec.~\ref{sec:rmf_res}, for a singular DoS the width can be a function of \(E\) and tend to zero, setting the limitation of the numerical study.

To analyze whether the eigenstates are localized or extended, we compute the box-counted participation number~\cite{evers2008anderson, wegner1980inverse, castellani1986multifractal, romer2010critical}
\begin{align}
    \label{eq:pn}
    \mathrm{PN}(E, \lambda)^{-1} = \sum_{\text{box}_j} \left(\sum_{i \in \text{box}_j} \abs{\psi_{E,i}}^2 \right)^{2} 
\end{align}
for the eigenstates inside the window of a target energy \(E\).
Here \(\lambda \equiv l/L\), where \(l\) is the length of each box of volume \(l^d\) and a divisor of \(L\).
Next we extract the scaling exponent of the PN, defined as
\begin{gather}
    \label{eq:ttau-def}
    \ttau = -\ln \apn  / \ln \lambda,
\end{gather}
where \( \langle \dots \rangle\) denotes the average over both disorder realizations and the eigenstates in the energy window around \(E\).
Note that for \(l = 1\) we get the conventional definition of PN without box-counting.
Finally the exponent \(\tau =\lim_{\lambda \to 0} \lim_{L \to \infty} \ttau\) describes the localization properties of the states around energy \(E\):
\begin{align}
    \label{eq:tau}
    \tau = 
    \begin{cases}
        0, \quad & \text{localized}\\
        d_c\quad (0 < d_c < d), \quad & \text{critical}\\
        d, \quad & \text{extended.}\\
    \end{cases}
\end{align}
The advantage of box-counting is that by fixing \(\lambda\) to a small value while increasing \(L\), we eliminate the irrelevant scaling effect with respect to \(\lambda\)~\cite{romer2010critical}.
Unless otherwise specified in the manuscript, \(\lambda = 0.04\) in all computations.

\section{Random magnetic flux disorder}
\label{sec:rmf}

In a tight binding model, the magnetic flux can be modeled by Peierls substitution, e.g. by adding phases to the hopping matrix elements (off-diagonal elements of the Hamiltonian)~\cite{peierls1933zur}.
The magnetic flux of a plaquette is proportional to the total phase change of the hoppings enclosing it.
One simple case is, for example, a square lattice with a constant magnetic flux, which gives rise to the famous Hofstadter butterfly spectrum~\cite{hofstadter1976energy}.
Here we are interested in another possibility -- random magnetic flux (RMF), which corresponds to assigning random phases to all the hoppings in a lattice, that is, 
\begin{align}
    t_{\mu m, \mu' n} \to t_{\mu m, \mu' n}\exp(i\phi_{\mu m, \mu' n})
\end{align}
where \(t_{\mu m, \mu' n} = \mel{\mu m}{\mh}{\mu' n} \) with (\(m \neq n\) or \(\mu \neq \mu'\)) and \( \mu, \mu' \in \{p, f\} \) and \(\phi_{\mu m, \mu' n}\) is a random phase with a box distribution \([-W/2, W/2]\), and
\(\phi_{\mu' n, \mu m} = -\phi_{\mu m, \mu' n}\) to preserve Hermiticity. 

\begin{figure}
	\includegraphics[width=0.5\columnwidth]{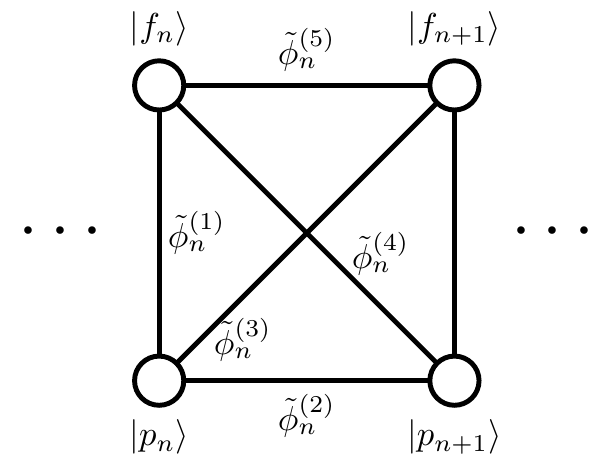}
	\caption{
        Definition of the random phase variables \(\tphi^{(j)}_n\), Eq.~\eqref{eq:rmf_hop}, attached to the hopping amplitudes in the \(n\)th unit cell of the 1D ABF \(\mh\)~(\ref{eq:fe}, \ref{eq:fe_1d_hop1}-\ref{eq:fe_1d_hop2}).
        Circles denote the lattice sites, and lines indicate the hoppings. 
        The three dots on the right and left indicate that this pattern repeats with periodic boundary condition.
    }
	\label{fig:fig2}
\end{figure}

In our work, we are interested in the case of infinitesimal \(W\), thus we can expand
\begin{align}
    t_{\mu m, \mu' n} \to t_{\mu m, \mu' n} + Wit_{\mu m, \mu' n}\tphi_{\mu m,\mu' n}
\end{align}
where \(\tphi_{\mu m,\mu' n}\) is scaled-out random phase with box distribution \([-1/2, 1/2]\).
The perturbation becomes additive and infinitesimal, therefore we are able to use Eq.~\eqref{eq:sf} to derive the scale-free model.
In \(d = 1\) the scale-free model \(\mhrmf\) is given by
\begin{widetext}
    \begin{align}
        \label{eq:rmf_sf}
        \mhrmf = \sum_n \tilde{t}_{n, n+1} \ketbra{a, n}{a, n+1} + \tilde{t}_{n, n+2}\ketbra{a, n}{a, n+2} + h.c.
    \end{align}
    \begin{subequations}
        \label{eq:rmf_hop}
        \begin{align}
            \begin{split}
                \tilde{t}_{n, n+1} {} &= 2 i\big(
                \tphi^{(1)}_{n} \cos^{4}\Th
                -\tphi^{(1)}_{n} \sin^{4}\Th 
                +\tphi^{(2)}_{n-1} \sin^{2}\Th\cos^{2}\Th
                +\tphi^{(2)}_{n} \sin^{2}\theta\cos^{2}\theta
                -\tphi^{(3)}_{n-1} \sin^{2}\theta\cos^{2}\theta \\
                &+\tphi^{(3)}_{n} \sin^{4}\theta
                +\tphi^{(4)}_{n-1} \cos^{4}\Th 
                -\tphi^{(4)}_{n} \sin^{2}\theta\cos^{2}\theta 
                -\tphi^{(5)}_{n-1} \cos^{4}\theta
                -\tphi^{(5)}_{n} \sin^{4}\theta
                \big) \sin^{2}{\theta} \cos^{2}{\theta}
            \end{split}
            \label{eq:rmf_nnhop} \\
            \tilde{t}_{n, n+2} &= 2 i \big(\tphi^{(2)}_{n} - \tphi^{(3)}_{n} - \tphi^{(4)}_{n} + \tphi^{(5)}_{n}\big) \sin^{4}\theta \cos^{4}\theta
            \label{eq:rmf_nnnhop}
        \end{align}
    \end{subequations}
\end{widetext}
This scale-free model describes a chain with no onsite potential and random, purely imaginary nearest, \(\tilde{t}_{n, n+1}\), and next nearest, \(\tilde{t}_{n, n+2}\), neighbor hoppings, with zero mean.
There are \(5\) hoppings for the \(n-\)th unit cell: \(1\) intra-cell and \(4\) inter-cell.
Correspondingly, there are \(5\) independent random phase variables \(\tphi^{(j)}_{n} \in [-1/2, 1/2]\) per unit cell, \(1 \leq j \leq 5\), as shown in Fig.~\ref{fig:fig2}.

Let us first consider a special case, \(\tilde{t}_{n, n + 2} = 0\) or \(\tphi^{(2)}_n + \tphi^{(5)}_n = \tphi^{(3)}_n + \tphi^{(4)}_n\) (the bracket in Eq.~\eqref{eq:rmf_nnnhop} vanishes).
In this case, the \(\mhrmf\) reduces to a linear chain with real random hoppings correlated over the neighboring unit cells~\footnote{
    We note that because the distributions of \(\tilde{t}\)'s are continuous, the probability to get an exact zero is zero.
    This implies that the very rare realisations where the chain is cut in two by a zero hopping are irrelevant.
}.
The energy spectrum is symmetric around \(E = 0\) due to chiral symmetry, with an \(E = 0\) state for odd system sizes.
The localization property of the eigenstates of that model is summarized in Sec.~\ref{sec:intro}.

For the general case, \(\tilde{t}_{n, n + 2} \neq 0\), the model appears to be different from the special well-known case, and does not have a chiral symmetry anymore because of the next-nearest neighbor hoppings.
However, now the particle-hole symmetry ensures that the spectrum is still symmetric around \(E = 0\).
To see this, from Eq.(\ref{eq:rmf_nnhop},\ref{eq:rmf_nnnhop}), we observe that the Hamiltonian is purely imaginary in matrix form.
Therefore, it is straightforward to check that the particle-hole transformation \(c^\dagger \to c\) keeps \(\mhrmf\) unchanged while flipping the sign of the energy in the eigenequation.
Similarly to the chiral case, if the number of lattice sites is odd, then there is an eigenstate precisely at \(E = 0\) (the flatband energy in the original model).

Despite the differences in symmetries (and hence the eigenstates) between the general and the special cases, in both cases the DoS diverges at \(E=0\) and the \(E=0\) eigenstates are sub-exponentially localized, as we show below.
For the n.n random hopping model, the mean DoS and the localization length are directly related by the Thouless formula~\cite{thouless1972dos}.

\subsection{Numerical results}
\label{sec:rmf_res}

\subsubsection{\texorpdfstring{\(d=1\)}{d=1}}

\begin{figure}
\centering
	\includegraphics[width=0.99\columnwidth, height = 0.48\columnwidth]{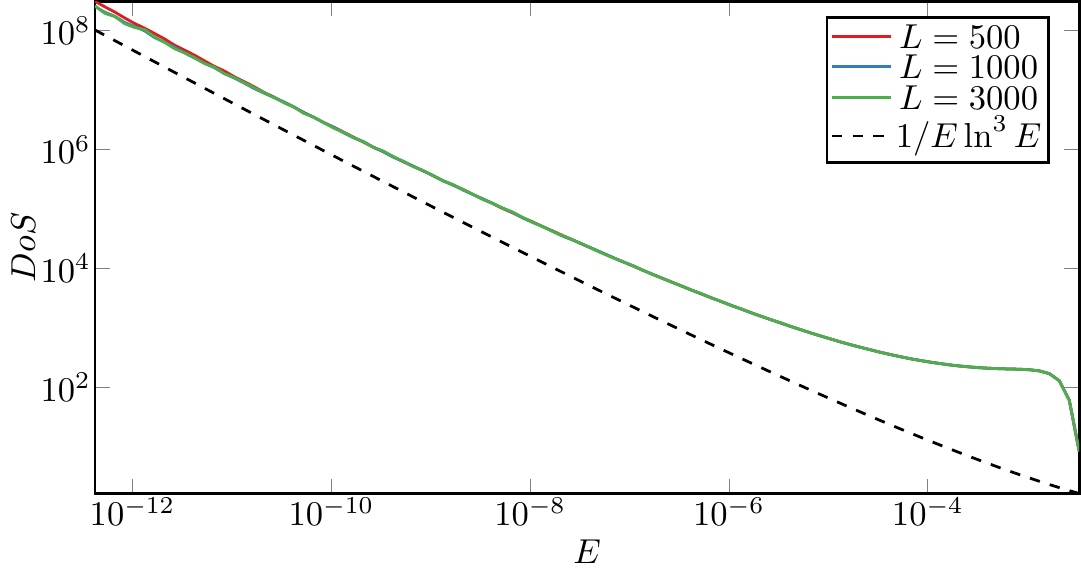}
	\caption{
	    Computed DoS (solid) for \(\theta=\pi/4\) and theoretical asymptotic behavior (dashed) of DoS for the special case.
    }
	\label{fig:fig3}
\end{figure}

We consider \(\theta=\pi/4\), which gives the strongest hopping, unless otherwise indicated.
We first compute the average DoS of the general model~\eqref{eq:rmf_sf} with next-nearest neighbor hoppings using exact diagonalization: it diverges for \(E\to 0\) just like in the random hopping model.
In Fig.~\ref{fig:fig3}, we show the comparison of our numerically computed DoS computation with the analytical prediction for the DoS of the random hopping chain. 
Both DoS show the same scaling \(E\to 0\), suggesting that both our random flux model and the random hopping chain have similar localization properties, despite having different symmetries.

We then study the \(\apn\), as given in Eq.~\eqref{eq:pn}, for the eigenstates of the scale-free Hamiltonian of \(d = 1\) ABF with RMF disorder, for nonnegative eigenenergies \(E\). 
As shown in Fig.~\hyperlink{subfig:fig4a}{4a}, the \(\apn\) for \(E\neq 0\) does not scale with \(L\), and the corresponding \(\tau = 0\), thus we conclude that all the corresponding eigenstates are localized. 
We also plot the \(\apn\) for the eigenstate at \(E = 0\), which was computed for odd number of sites, \(L = 301\) and was only averaged over disorder realizations.~\footnote{We find/confirm that the \(\apn\) of the \(E = 0\) eigenstate does not scale with \(L\) by computing it for \(L = 501, 701, 1001\).}
The \(\apn\) close to \(E = 0\) is resolved in the inset of Fig.~\hyperlink{subfig:fig4a}{4a} where the \(\apn\) vs \(E\) is plotted on the logarithmic scale.
The \(\apn(E\to 0)\) weakly scale with \(L\), but the finite size scaling result of \(\ttau\) in Fig.~\hyperlink{subfig:fig4b}{4b} indicates \(\tau(E\to0) = 0\) and these eigenstates are also localized.
The results for other values \(\theta\) are qualitatively similar and are not shown here.

Next we focus on the special behavior around \(E=0\), that we discussed above: both density of states (DoS) and localization length \(\xi\) diverge at \(E=0\) and there is a special eigenstate at \(E=0\) for odd system sizes \(L\).
In order to study the vicinity of the band center \(E = 0\), we use the transfer matrix method~\cite{mackinnon1981one, pichard1981finite, mackinnon1983the, slevin2014critical}. 
The computed  \(\xi\) diverges as \(\ln E\) for \(E\to 0\) as shown in Fig.~\hyperlink{subfig:fig5a}{5a}.
This contradicts the prediction based on the scaling of \(\apn\) with \(L\), that the \(E=0\) state is localized.
An explanation of this apparent discrepancy is that \(\xi\) is a property of an asymptotic tail of the eigenstate, while \(\apn\) measures the core of the eigenstate.
If the single parameter scaling holds~\cite{kramer1993localization}, \(\xi\) is the single parameter that determines the length scale of the system, which is clearly not true in our case.
Similar logarithmic divergence of \(\xi\) is also observed in the chain with hopping disorder near the band center.
This is the anomalous localization, or \textit{freezing} \cite{carpentier2001glass, motrunich2002particle, evers2008anderson}, and can be distinguished from exponential localization by examining \(\tilde{\tau}(q)\), where the final exponent in Eq.~\eqref{eq:pn} is changed from \(2\) to \(q\).

%
%
\begin{figure}
    \hypertarget{subfig:fig4a}{}
    \hypertarget{subfig:fig4b}{}  
    \includegraphics[width=\columnwidth]{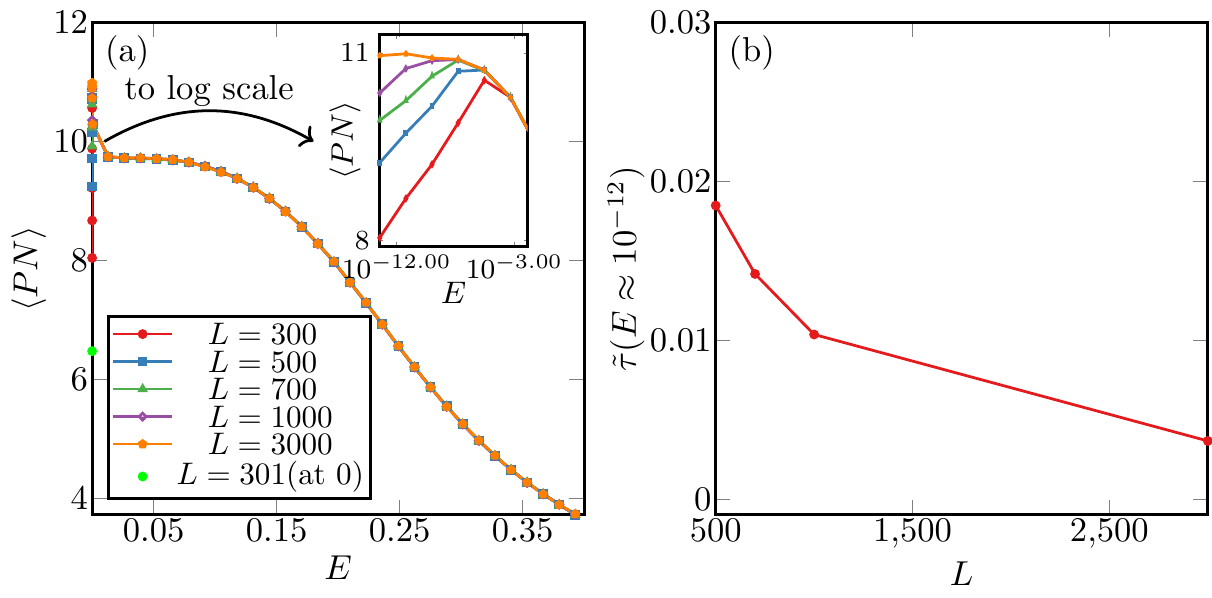}\hfill
    \caption{
        (a) The average participation number \(\apn\), with box size \(l = 1\), for the \(d = 1\) \(\mhrmf\) vs \(E\).
        The inset shows the \(\apn\) near \(E = 0\), on a logarithmic energy scale. 
        (b) The exponent \(\ttau(L)\) at \(E \approx 10^{-12}\), the leftmost points in the inset of (a): \(\lim_{L \to \infty }\ttau = 0\) implying all these eigenstates are also localized.
    }
    \label{fig:fig4}
\end{figure}

Further we attempt to quantify the localization of the \(E=0\) eigenstate: we expect it to have the same sub-exponential, \(\exp(-\sqrt{r})\), localization as in the random hopping model.
This is a challenging problem: the transfer matrix method has to be modified since the localization length is infinite for \(E=0\).
Furthermore, the results shown in Fig.~\ref{fig:fig4} and Fig.~\hyperlink{subfig:fig5a}{5a} indicate that even tiny deviations away from \(E=0\) give finite and strongly varying values of \(\xi\).
For instance, the \(\xi(E = 10^{-100})\) and \(\xi(E = 10^{-10})\) differ by an order of magnitude.
Exact diagonalization faces a number of problems:
(a) relatively large \(L\) are required to capture the decay of the \(E=0\) eigenstate
(b) the mean DoS diverges at \(E=0\), the level spacing around \(E = 0\) quickly becomes so small with growing \(L\) that the exact diagonalization eigenstate at \(E=0\) get contaminated by states with eigenenergies indistinguishable from zero within machine precision.
Therefore, we solve the linear system \(H\ket{\psi} = 0\) directly~\cite{xiong2001power} using arbitrary precision arithmetic.
We increase the precision until the desired convergence of the \(\psi_n\) is reached for all sites \(n\), e.g. the amplitudes \(\psi_n\) stop changing with increasing precision.
This algorithm allows us to obtain the eigenstate at \(E = 0\).

\begin{figure}
    \hypertarget{subfig:fig5a}{}
    \hypertarget{subfig:fig5b}{}    
    \includegraphics[width=0.475\columnwidth]{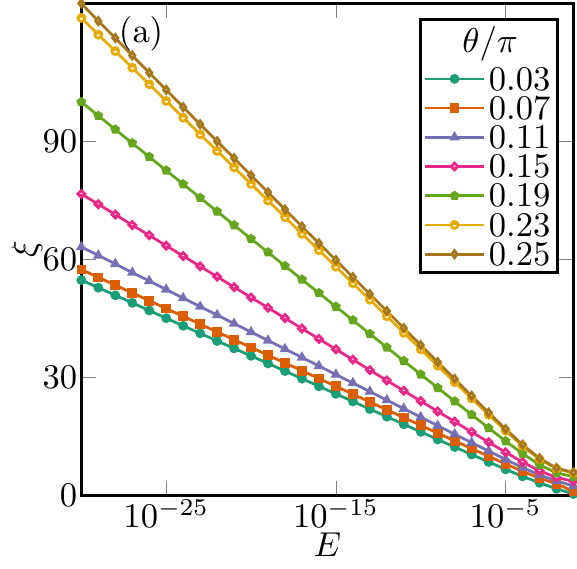}\hfill
    \includegraphics[width=0.525\columnwidth]{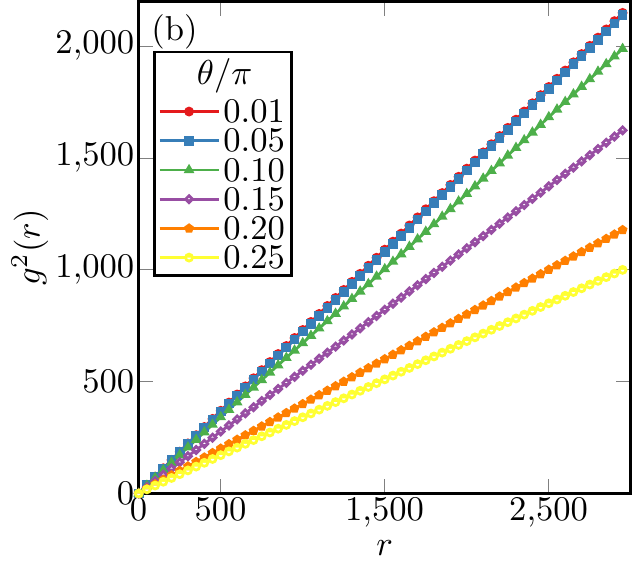}
    \caption{
        Random magnetic field model \(\mhrmf\) in \(d=1\).
        \textbf{(a)} \(E \neq 0\): Localization length \(\xi\) vs \(E\) as a function of the energy, for various angle \(\Th\).
        A clear \(\ln E\)-divergence of the localization length is observed.
        \textbf{(b)} \(E = 0\): Squared average logarithmic decay \(g^2(r)\)~\eqref{eq:gr} vs distance \(r\) computed for the \(E = 0\) eigenstate. 
        The linear behavior signals sub-exponential localization.
    }
    \label{fig:fig5}
\end{figure}
%
%
\begin{figure}
    \hypertarget{subfig:fig6a}{}
    \hypertarget{subfig:fig6b}{}
    \hypertarget{subfig:fig6c}{}
    \hypertarget{subfig:fig6d}{}
    \hypertarget{subfig:fig6e}{}
    \hypertarget{subfig:fig6f}{}

    \includegraphics[width=\columnwidth]{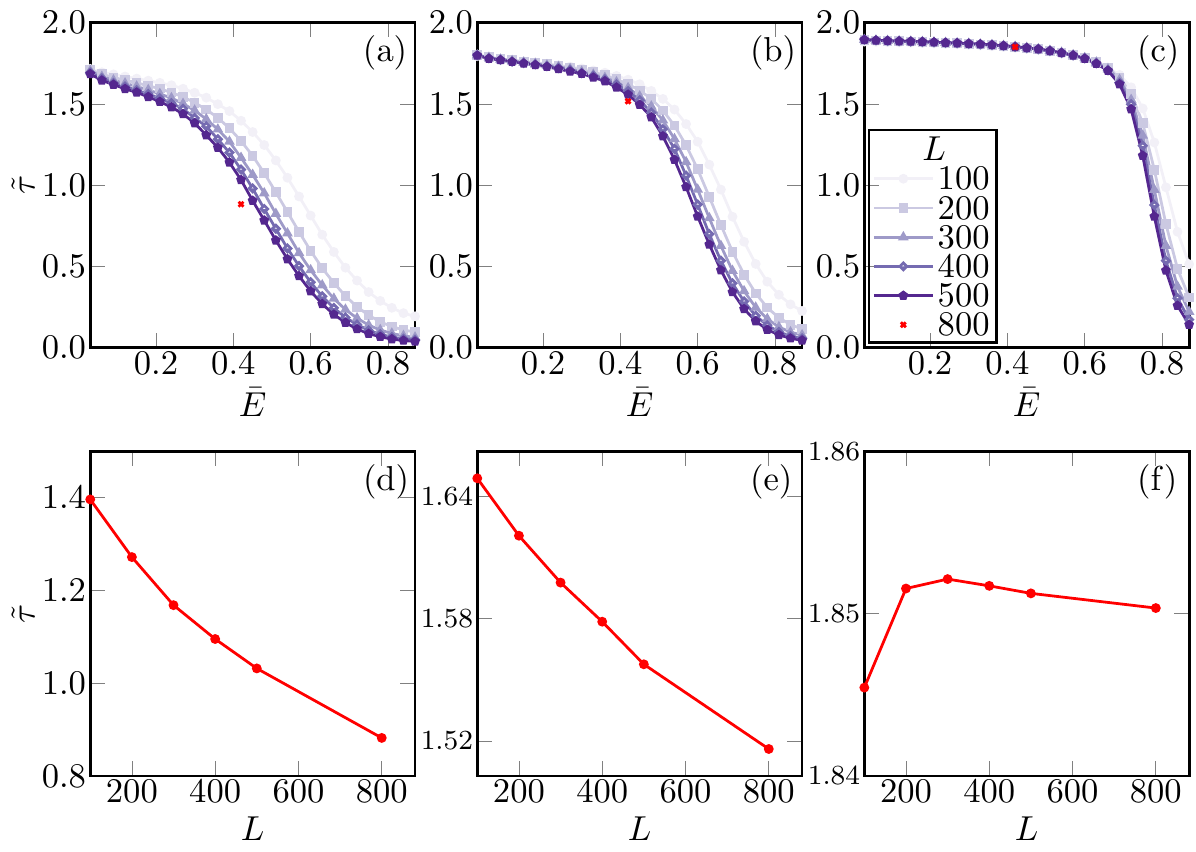}
    \caption{
        The 2D random magnetic field model \(\mhrmf\) for linear sizes \(L = 100, 200, 300, 400, 500, 800\).
        \(\apn\) was averaged over \(7500, 3400, 500, 200, 100, 100\) disorder realizations.
        Data for \(L=800\) were computed for \(\bE = 0.42\) only.
        \(\ttau\) is computed with fixed \(\lambda = l/L = 0.04\) (See Eq.~\eqref{eq:ttau-def}).
        Top row: \(\ttau\) vs. \(\bE\) (\(0.03 \leq \bE \leq 0.9\)).
        Bottom row: \(\ttau\) vs. \(L\), at \(\bE = 0.42\).
        Left to right: \(\theta = 0.01\pi, 0.125\pi, 0.25\pi\) for (a,d), (b,e) and (c,f) respectively.
    }
    \label{fig:fig6}
\end{figure}
%
%
\begin{figure}
    \hypertarget{subfig:fig7a}{}
    \hypertarget{subfig:fig7b}{}
    
    \includegraphics[width=\columnwidth]{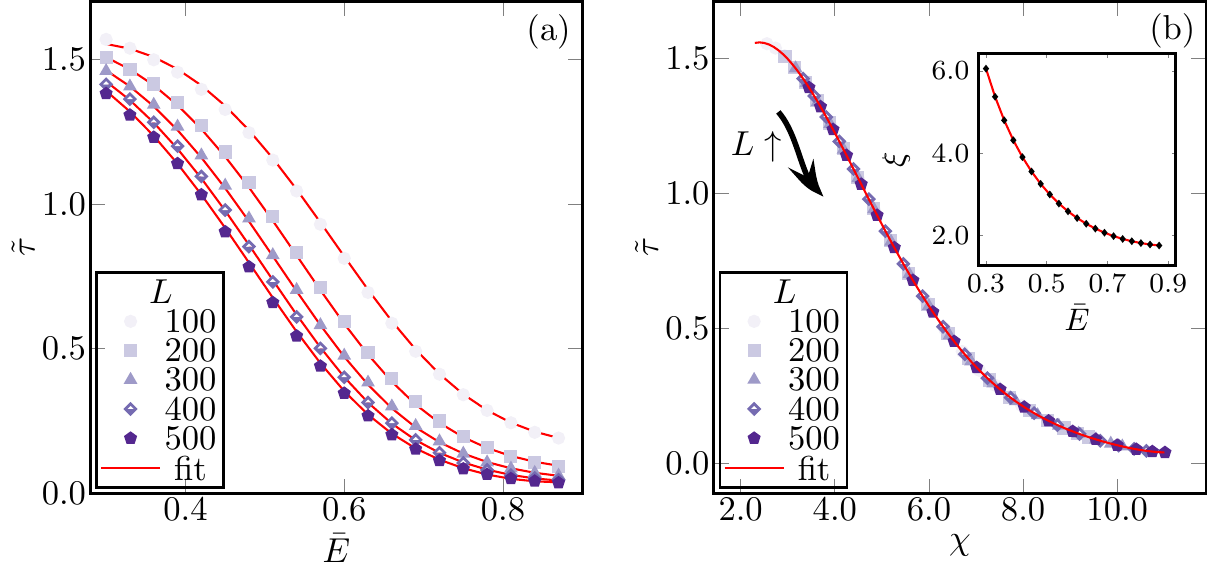}
    \caption{
        The 2D random magnetic field model \(\mhrmf\).
        Finite size scaling analysis of \(\ttau\) at \(\Th =0.01\pi\) in the rescaled energy range \(0.3 < \bE < 0.9\) of the data in Fig.~\protect\hyperlink{subfig:fig6a}{6a}.
        (a) Nonlinear least squares fitting of \(\ttau\). 
        (b) Collapse of \(\ttau\) as a function of \(\chi = \ln L / \ln \xi\). 
        (Inset) Fit for the localization length \(\xi\) (up to a constant prefactor).
    }
	\label{fig:fig7}
\end{figure}

The anomalous localization of the \(E=0\) eigenstate are analyzed using the average logarithmic decay \(g(r)\)~\cite{inui1994unusual, krishna2020beyond}, defined as,
\begin{align}\label{eq:gr}
    g(r) = \langle \abs{\ln \abs{\psi_{E, n_{\mathrm{max}}}}} - \abs{\ln \abs{\psi_{E, n_{\mathrm{max}+r}}}} \rangle
\end{align}
where \(n_\mathrm{max}\) is the position of the maximum (in absolute value) of the \(E=0\) eigenstate, and \(\langle \dots \rangle\) is the average over disorder realizations.
For an exponentially localized state, \(g(r)\) is asymptotically linear, while for an extended state, \(g(r)\) is asymptotically zero.
The \(g(r)\) computed using the linear solver is shown in Fig.~\hyperlink{subfig:fig5b}{5b}: we plot \(g^2(r)\) instead of \(g(r)\) to capture the expected sub-exponential localization.
We see clearly that the \(E = 0\) eigenstate is sub-exponentially localized.
From these numerical results, we see that the localization properties of \(d=1\) ABF with RMF disorder are qualitatively similar to that of the chain with hopping disorder:
there is a divergent DoS at \(E=0\) and a sub-exponentially localized \(E=0\) eigenstate.

\subsubsection{\texorpdfstring{\(d = 2\)}{2D}}

We have also looked at the \(d=2\) ABF model with random magnetic flux using the same approaches and methods as in \(d=1\).
For convenience, we introduce the rescaled energy \(\bE \in [-1, 1]\). 
Our numerical results for nonzero energies are summarized in Fig.~\ref{fig:fig6}, which shows the scaling of the exponent \(\ttau\) vs. linear system size \(L\) (\(L^2\) unit cells), for several values of \(\theta\).
The trend in scaling of \(\ttau\) vs \(L\) for small \(\theta\), observed in Figs.~\hyperlink{subfig:fig6a}{6a} and~\hyperlink{subfig:fig6b}{6b} suggests that all the eigenstates are localized for nonzero \(\bE\), as \(\ttau\) is reducing with increasing system size. 
To confirm that \(\ttau\to 0\) in the \(L \to \infty\) limit, we perform a finite size scaling analysis for the case \(\Th = 0.01\pi\) and show the results and the extracted localization lengths in Fig.~\hyperlink{subfig:fig7}{7}. 
These results support localization away from \(E=0\).
We note however that extremely large system sizes, well beyond the accessible ones, would be needed to see \(\ttau\approx 0\), especially for \(\theta\) around \(0.25\pi\).
The details of the fitting procedure are given in Appendix~\ref{app:fss_rmf}.

At \(\Th = 0.25\pi\), the case of the strongest hopping in the clean Hamiltonian, the \(\ttau\) near \(\bE = 0\) seem to lay on top of each other for different linear sizes \(L\) (Fig.~\hyperlink{subfig:fig6c}{6c}), suggesting critical eigenstates according to Eq.~\eqref{eq:tau}.
However, the likely scenario is that all the eigenstates are also localized, but the localization length \(\xi\) becomes much larger than the considered linear system sizes \(L\), e.g. \(L \ll \xi\), and huge system sizes are required to observe that at small \(\bE\).
This is supported by the finite size scaling and larger values of \(\bE\) and the scaling of \(\ttau\) for smaller values of \(\Th\).

\begin{figure}
    \hypertarget{subfig:fig8a}{}
    \hypertarget{subfig:fig8b}{}
    \includegraphics[width=\columnwidth]{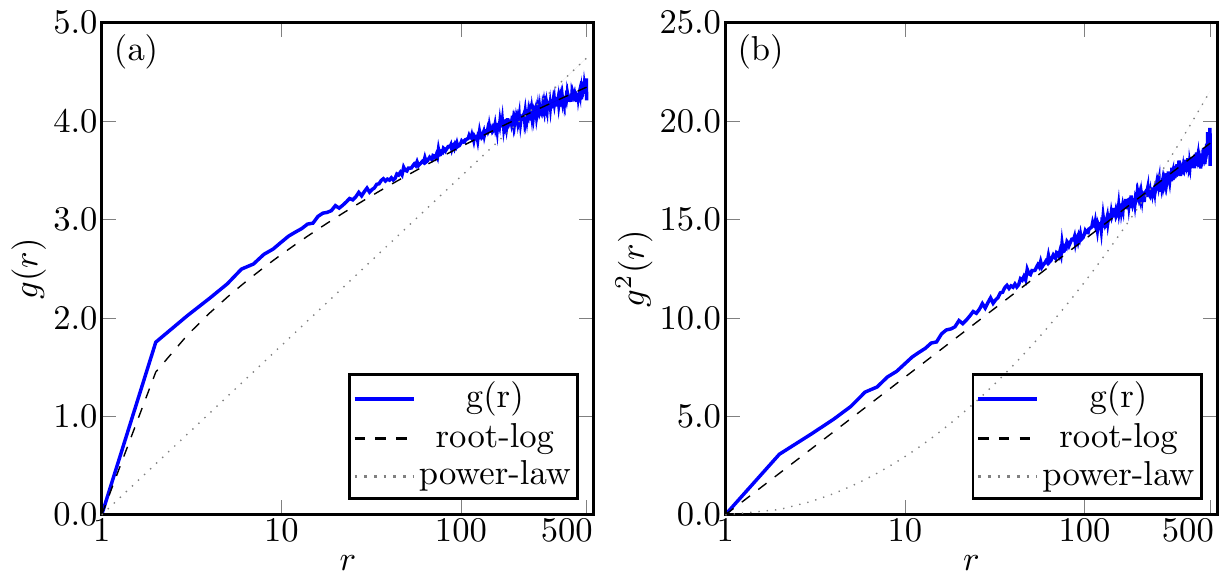}
    \caption{
        The average logarithmic decay, \textbf{(a)} \(g(r)\) in \textbf{(b)} \(g^2(r)\), computed for the \(\bE = 0\) eigenstate of the \(d = 2\) RMF model with \(L=501\) and \(\theta = 0.19\pi\) and averaged over \(2000\) realizations. 
        The resulting curve appears to agree better with the \((\gamma_{\textrm{Ziman}}\ln |r|)^{1/2}\) fit (\(\gamma_{\textrm{Ziman}} = 1.74\)) (dashed black) predicted analytically by Ziman~\cite{ziman1982localziation} rather than the \(\gamma_{\textrm{power}} \ln |r|\) (\(\gamma_{\textrm{power}} = 0.75\)) power-law fit (dotted grey).
        }
	\label{fig:fig8}
\end{figure}

Similarly to the \(d=1\) model, the \(d=2\) model features a special \(\bE=0\) eigenstate, which is enforced by the particle-hole symmetry and also anomalously localized.
We computed the average logarithmic decay \(g(r)\) of the \(\bE = 0\) eigenstate for the system size \(L = 501\) and \(\theta = 0.19\pi\) using the same method as \(d = 1\).
The result is shown in Fig.~\ref{fig:fig8}.
We fit the \(g(r)\), which is the logarithm of the wavefunction amplitude, with two different curves by varying the exponent \(\gamma\): \((\gamma\ln |r|)^{1/2}\) -- an analytical prediction by Ziman~\cite{ziman1982localziation} -- and the power-law decay \(\ln |r|\) -- the numerical result suggested by Xiong and Evangelou~\cite{xiong2001power} for the square lattice with hopping disorder.
The fitting results (\(\gamma_{\textrm{Ziman}} = 1.74\) and \(\gamma_{\textrm{power}} = 0.75\))
in Fig.~\hyperlink{subfig:fig8b}{8b}, with  suggest that \(g(r)\) shows a slower than the power-law decay, rather the \(g^2(r)\) is following the root-log decay.

\section{Spin-orbit coupling disorder}
\label{sec:soc}

Here we consider the case of weak SOC disorder.
We first construct an uncoupled spin-\(\frac{1}{2}\) single particle clean ABF Hamiltonian \(\mh_s\) by taking two copies of the identical spinless Hamiltonian \(\mh\).
To do this, we double the Hilbert space to include spin degrees of freedom, \(\ket{p, n} \to \ket{p, n, \sigma} \):
\begin{align}
    \label{eq:soc_uncoupled}
    \mh_{s} 
    &= \sum_{n,n', \mu, \sigma}
    H_{\mu n, \mu' n'}\ket{\mu, n, \sigma}
    \bra{\mu', n', \sigma}
\end{align}
where \(\mu, \mu' \in \{p, f\}\), \(\sigma\in \{\uparrow, \downarrow\}\).
For the off-diagonal elements of \(\mh\) (i.e. \(n \neq n'\) or \(\mu \neq \mu'\)), let \(t_{\mu n, \mu' n'} = H_{\mu n, \mu' n'}\).
The SOC is added through the spin-flip hoppings, with the following constraint ensuring the absence of external magnetic fields and the persistence of time reversal symmetry~\cite{ando1989numerical}:
\begin{align}
    \label{eq:soc_hopping}
    t_{\mu n, \mu' n'} I_{2\times2}
    \to t_{\mu n, \mu' n'} X_{\mu n, \mu' n'}
\end{align}
where
\begin{align}
\label{eq:soc_constraint}
    X_{i,j} = 
    \begin{pmatrix}
        \phantom{+}e^{i\alpha_{i,j}} \cos \beta_{i,j} & \phantom{+}e^{i\gamma_{i,j}}\sin \beta_{i,j} \\
        - e^{-i\gamma_{i,j}} \sin \beta_{i,j} & \phantom{+}e^{-i\alpha_{i,j}} \cos \beta_{i,j} 
    \end{pmatrix}
\end{align}
for \(i = (\mu n)\) and \(j = (\mu' n')\).
\(\beta_{i,j}\) are random variables with box distribution \([-W/2, W/2]\) and \(\gamma_{i,j}\) and \(\alpha_{i,j}\) are uniform random variables in \([-\pi, \pi]\).

In the weak coupling limit, \(\beta_{i,j} \to 0\) and we approximate \(\cos \beta_{i,j} \approx 1\) and \(\sin \beta_{i,j} \approx \beta_{i,j}\).
We also set \(\alpha_{i,j} = 0\) so that the perturbation is additive in the weak \(W\) limit. 
Then, we apply Eq.~\eqref{eq:sf} to derive the scale-free model, \(\mhsoc\) for \(d = 1\):

\begin{widetext}
    \begin{align}
        \label{eq:soc_sf}
        \mhsoc &= \sum_n (\tilde{T}_{n, n+1})_{\sigma \sigma'} \ket{a, n, \sigma} \bra{a, n+1, \sigma'} + (\tilde{T}_{n, n+2})_{\sigma \sigma'}\ket{a, n, \sigma} \bra{a, n+2, \sigma'} + h.c.\\
        \label{eq:soc_hop_mat}
        \tilde{T}_{n, m} &= 
        \begin{pmatrix}
            0 & \tilde{t}_{n, m}\\
             -\tilde{t}^*_{n, m} & 0
        \end{pmatrix}
    \end{align}
    \begin{subequations}
    \label{eq:soc_hop}
        \begin{align}
            \begin{split}
                \tilde{t}_{n, n+1} &= 2(
                 \om^{(1)}_{n} \cos^4 \Th 
                -\om^{(1)}_{n} \sin^4 \Th
                +\om^{(2)}_{n-1} \sin^2 \Th \cos^2 \Th 
                +\om^{(2)}_{n} \sin^2 \Th \cos^2 \Th
                -\om^{(3)}_{n-1}\sin^2 \Th \cos^2 \Th \\
                &+\om^{(3)}_{n} \sin^4 \Th
                +\om^{(4)}_{n-1}\cos^4 \Th
                -\om^{(4)}_{n}\sin^2\Th\cos^2\Th
                -\om^{(5)}_{n-1}\cos^4 \Th
                -\om^{(5)}_{n} \sin^4 \Th)\sin^2 \Th \cos^2 \Th
            \end{split} \\
            \tilde{t}_{n, n+2} &= 2(\omega^{(2)}_{n}-\omega^{(3)}_{n}-\omega^{(4)}_{n}+\omega^{(5)}_{n})\sin ^4 \theta \cos^4 \theta
        \end{align}
    \end{subequations}
\end{widetext}

Here, \(\tilde{t}_{n, m}\) are correlated spin-flip hoppings with zero mean, where \(\omega^{(j)}_{n} = \beta^{(j)}_{n}\exp(i\gamma^{(j)}_{n})\) and, \( 1 \leq j \leq 5\) is an index for 5 hoppings in a \(n\)th unit cell.
The hoppings in Eq.~\eqref{eq:soc_hop} and Eq.~\eqref{eq:rmf_hop} are similar, except \(\omega^{(j)}_n\) are complex random variables, while the hoppings in Eq.~\eqref{eq:rmf_hop} are purely imaginary.
We also note that \(\tilde{T}_{n,m}\)~\eqref{eq:soc_hop_mat} still satisfies the constraint~\eqref{eq:soc_constraint}, implying that the \(\mhsoc\) also belongs to the symplectic class~\cite{haake1991quantum}.
The projected scale-free model \(\mhsoc\) features an additional chiral symmetry between the spin-up and spin-down sublattices, which is destroyed for finite values of \(W\), since \(\tilde{T}_{n,m}\) only contains spin-flip hoppings, and no spin preserving hoppings. 
Therefore, for odd lattice sizes Kramer's degeneracy~\cite{kramers1930theorie} ensures that there is a double degenerate \(E = 0\) state, which can be decomposed into purely spin-up and spin-down states.
However, unlike the RMF disorder case considered before, we found no evidence of divergent DoS at \(E = 0\) or localization length in this model.

Similar to the symplectic class Hamiltonian discussed in the introduction Sec.~\ref{sec:intro}, we demonstrate below that the MIT occurs also in \(d = 2\) ABF SOC disordered systems in the weak disorder limit \(W\to0\).

\subsection{Numerical results}
\label{sec:soc_res}

We find that in \(d=1\) all the eigenstates are localized for the weak SOC disorder case, \(\mhsoc\)~\eqref{eq:soc_sf}.
Although SOC disorder is similar to the RMF disorder in that both cases give off-diagonal disorder, we found that all states are localized.
Therefore we focus on the \(d = 2\) case.

Due to the complexity of the scale-free Hamiltonian with the SOC disorder in \(d=2\), we derive and diagonalize it numerically.
We again define \(\bE\) to be the rescaled energy so that the spectrum extends from \(-1\) to \(1\). 
Fig.~\hyperlink{subfig:fig9a}{9a} shows the exponent \(\ttau\) of the \(\apn\) of the eigenstates, Eq.~\eqref{eq:ttau-def}, vs. \(\bar{E}\) at \(\theta = 0.01\pi\) computed for increasing system sizes \(L\).
We see that \(\ttau\) is approaching zero value (localized) at the band edge and value \(2\) (extended) near the center of the band.
Curves for different \(L\) cross, indicating a mobility edge: This transition is clearly seen in the inset of Fig.~\hyperlink{subfig:fig9a}{9a}.
The results for other \(\theta\) are qualitatively similar.

\begin{figure}
    \hypertarget{subfig:fig9a}{}
    \hypertarget{subfig:fig9b}{}
    \hypertarget{subfig:fig9c}{}
    \hypertarget{subfig:fig9d}{}
    
    \includegraphics[width = \columnwidth]{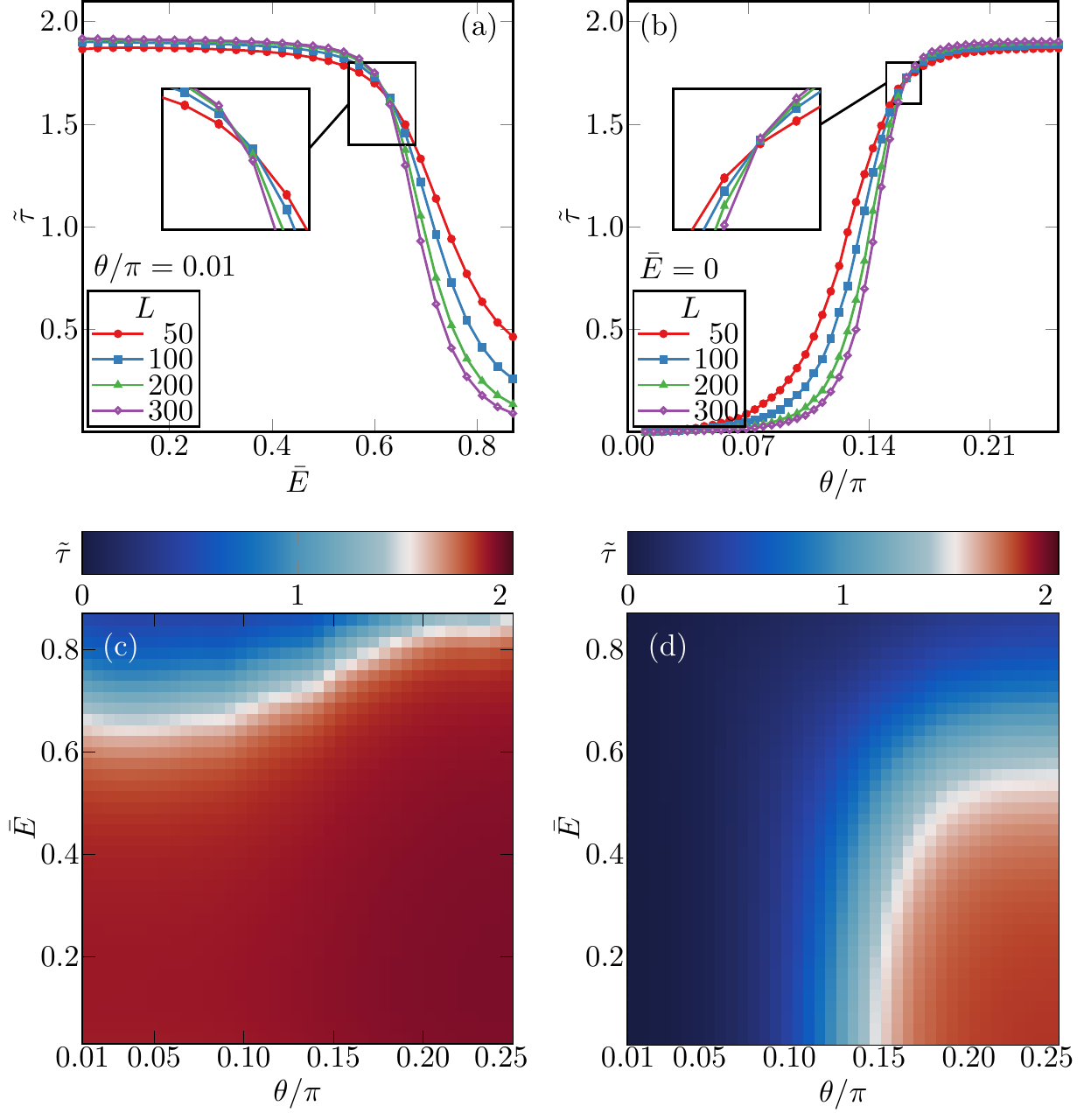}
    \caption{
	    The system size scaling of the exponent of the participation number \(\ttau\) vs. \(\bE\) and \(\Th\) 
	    for the pure SOC  and both SOCO (\(\delta =  2\)) is shown in panels (a) and (b), respectively. 
	    \textbf{(a)} \(\ttau(E, \Th/\pi = 0.01)\) for pure SOC;
	    \textbf{(b)}  \(\ttau(E = 0, \Th)\) for SOCO.
	    The system size consists of \(L^2\) unit cells, with \(L = 50, 100, 200, 300\).
	    \(\apn\) average was done over \(\approx 10^4\) states.
        \(\ttau\) is computed with fixed \(\lambda = l/L = 0.04\) (See Eq.~\eqref{eq:ttau-def}).
	    In both cases crossing points are present, signaling the phase transition.
	    The insets zoom into the vicinity of the crossing points.
	    \textbf{(c)} Phase diagram (based on the value of \(\ttau\)) vs. \(\Th\) and \(\bE\) for the SOC case, with \(L=50\).
	    The red region is a metal while the blue region is an insulator.
        The data in (a) correspond to the left border of the diagram. 
	    \textbf{(d)} Phase diagram as in (c), but for the SOCO case with \(L=50\).
        The data in (b) corresponds to the bottom border of the diagram, \(\bE = 0\).
    }
    \label{fig:fig9}
\end{figure}

\begin{figure}
    \hypertarget{subfig:fig10a}{}
    \hypertarget{subfig:fig10b}{}
    \hypertarget{subfig:fig10c}{}
    \includegraphics[width = \columnwidth]{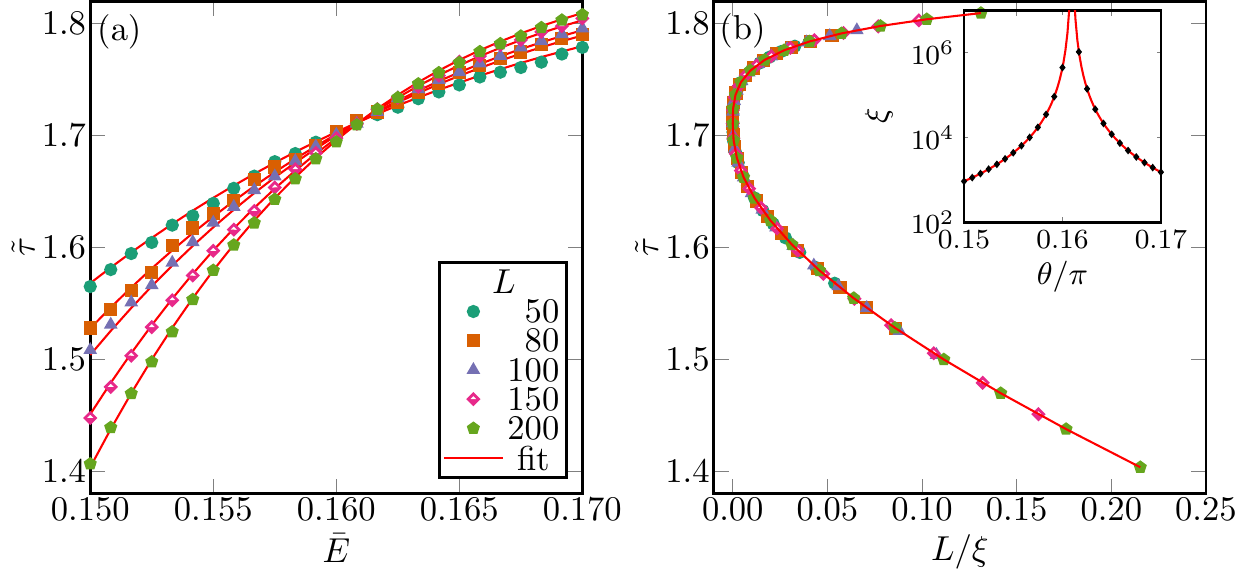}
    \caption{
        Finite size scaling analysis of the ABF with weak spin-orbit coupling and onsite potential disorder, with \(\delta=2\), (Sec.~\ref{app:fss_soco}).
        The FSS is done at \(E=0\) for the parameters used in Fig.~\protect\hyperlink{subfig:fig9b}{9b}.
        \textbf{(a)} Dashed line is the non-linear least squares fitting result of the scaling function \(F\) to the data points \(\ttau(\theta)\).
        \textbf{(b)}: \(F\) as a function of \(L/\xi\). 
        The data collapse shows that the one-parameter scaling hypothesis works well.
        \textbf{Inset of (b)}: Divergence of the localization length \(\xi\) near the critical point \(\theta_c/\pi = 0.1621 \pm 0.0005\).
    }
    \label{fig:fig10}
\end{figure}

We also consider the case where both SOC and the onsite potential disorders are present, which we refer to as \emph{SOCO} case: 
in addition to the perturbation in Eq.~\eqref{eq:soc_hopping} with strength \(W\) the onsite potential (diagonal) disorder with infinitesimal strength \(V\) term is added.
This is the type of perturbation which produces the \(\mathrm{SU(2)}\) model on a square lattice~\cite{asada2002anderson, asada2004numerical}, although we only consider infinitesimal strengths of the  onsite disorder. 
The onsite disorder is given by
\begin{align}
    \label{eq:onsite_perturbation}
    V\mdon = V\sum_{n,\sigma} \tilde{\varepsilon}_{n, \mu} \ketbra{\mu, n, \sigma}{\mu, n, \sigma}
\end{align}
where \(\tilde{\varepsilon}_{n, \mu}\) are random variables with box distribution in \([-1/2, 1/2]\), and \(\sigma \in \{\uparrow, \downarrow\}\), and \(\mu \in \{p, f\}\).
The effective projected Hamiltonian is given by
\begin{align}
    \mhsoco = \mhsoc + \delta \cdot \mhon
\end{align}
where \(\mhon\) is the scale-free model created from infinitesimal disorder Eq.~\eqref{eq:onsite_perturbation} using Eq.~\eqref{eq:sf}, and \(\delta = V/W\) the ratio of infinitesimal SOC disorder to onsite disorder. 
As explained earlier, the analytical expression for the scale-free model is too complex, and the Hamiltonian was instead evaluated numerically.

Unlike the random magnetic flux or SOC scale-free models, there is an additional scale remaining in the process of scaling out of \(W\) from the effective model~\eqref{eq:sf}. 
Thus, we no longer obtain a scale-free model but a model with a control parameter given by the ratio of the strengths of the two disorders, \(\delta\).
We set \(\delta = 2\) in our numerics, so that both disorders are of the same order.

Based on the results with no onsite disorder, Fig.~\hyperlink{subfig:fig9a}{9a}, we expect the onsite potential disorder, 
which tends to localize the system, to compete with the SOC generated hopping disorder controlled by the angle \(\theta\) in the projected model.
The exponent \(\ttau\) as a function of \(\theta\) at energy window around \(E = 0\) for the SOCO case is shown in Fig.~\hyperlink{subfig:fig9b}{9b}.
Since the curves for different \(L\) cross as for the pure SOC disordered case, the MIT is now induced by changing \(\theta\) (at \(E \approx 0\)).
For the fixed ratio \(\delta\) we see a transition at a critical angle \(\theta_c\) as shown in Fig.~\hyperlink{subfig:fig9b}{9b}.

Figures~\hyperlink{subfig:fig9c}{9c} (SOC model) and~\hyperlink{subfig:fig9d}{9d} (SOCO model) are the 2D color plots of \(\ttau\) vs \(\bE\) and \(\ttau\) vs \(\theta\).
Fig.~\hyperlink{subfig:fig9a}{9a} corresponds to the left border of the color plot in Fig.~\hyperlink{subfig:fig9c}{9c}, and Fig.~\hyperlink{subfig:fig9b}{9b} corresponds to the bottom border of the color plot in Fig.~\hyperlink{subfig:fig9d}{9d}.
The blue color corresponds to the localized phase and the red to the extended phase.
It is clear that onsite disorder in the SOCO model creates a MIT as a function of \(\theta\), Fig.\hyperlink{subfig:fig9d}{9d}, as compared to the extended bulk and localized edges of the spectrum in Fig.~\hyperlink{subfig:fig9c}{9c}.

In the vicinity of the critical angle \(\theta_c\) we conduct the finite size scaling analysis for the SOCO model as shown in Fig.~\ref{fig:fig10} (For details, see Appendix~\ref{app:fss_soco})
A universal scaling function \(\ttau = F(L/\xi)\) is assumed and expanded as a polynomial with unknown coefficients which are fitted using non-linear least squares fitting algorithms.
The single parameter scaling hypothesis works well as shown in Fig.~\hyperlink{subfig:fig10b}{10b}: all the data points collapse into a single curve \(F(L/\xi)\).
We also extract the localization length (correlation length) \(k\xi\) up to an arbitrary constant \(k\) from the fitting, in the inset of Fig.~\hyperlink{subfig:fig10b}{10b}.
The critical point is estimated as \(\theta_c/\pi = 0.1621 \pm 0.0005\) for this choice of ratio, \(\delta=2\).
The critical exponent \(\nu\), describing the divergence of the localization length \(\xi\) is \(\nu \approx 2.70 \pm 0.05\), in agreement with the critical exponent obtained for the \(\mathrm{SU}(2)\) model~\cite{asada2002anderson, asada2004numerical}, which belongs to the symplectic \(d=2\) class.

\section{Conclusions}
\label{sec:conclusion}

We studied the effect of the weak random magnetic flux disorder and weak spin orbit coupling plus weak onsite disorder on \(d = 1,2\) all bands flat lattice manifolds. 
The manifolds are constructed by local unitary transformations, and their reaction to weak disorder is described by effective scale-free models, which capture the non-perturbative effects of disorder on a chosen flatband.
We then have studied the localization properties of the scale-free models analytically and numerically.
The non-perturbative effect of weak disorder in all bands flat lattices results in complete destruction of compact localized states and counterintuitively can lead to a non-perturbative metal-insulator transition in 2D in the presence of spin-orbit coupling.
In such cases the reentrant behavior is likely observed upon further increase of disorder, e.g. localization at strong enough disorder.

Interestingly, for both types of disorder considered, the effective models inherit the symmetry of the weak perturbation (e.g. particle-hole symmetry for RMF disorder, and chiral symmetry for SOC disorder).
This leads to an unconventional, slower than exponential localization(freezing) precisely at the flatband energies.
Further investigations using multifractal analysis may reveal more facts about the anomalous localization, or freezing effect \cite{carpentier2001glass, motrunich2002particle}.

For nonzero energies, the effective models follow the universality of Anderson localization for the three Wigner-Dyson symmetry classes: 
for the random magnetic flux, we see localization in \(d = 1, 2\), while for the spin-orbit coupling, we see localization in \(d = 1\) and an energy dependent metal-insulator transition induced by the manifold angle parameter \(\Th\).
In the latter case the critical exponent is \(\nu = 2.70 \pm 0.05\), in agreement with the critical exponent previously reported for the square lattice with spin-orbit coupling~\cite{asada2002anderson, asada2004numerical}. 

Finally, we briefly discuss the case of finite strength disorder.
Generically strong enough disorder would localize all the eigenstates.
Projected models stop being valid and no longer describe the properties of the eigenstates once the disorder becomes comparable to the bandgap.
The particle-hole(chiral) symmetry of the RMF(SOC) model is destroyed,
as well as the unconventionally localized eigenstate at the flatband energy (corresponding to the \(E = 0\) state of the scale-free model).
Interesting phenomena might occur for disorder strength comparable to the bandgap, when the two perturbed flatbands hybridize.
In particular, we observed a nontrivial enhancement of localization length in a 1D model in this regime~\cite{cadez2021metalinsulator}. 
An \emph{inverse Anderson transition}, e.g. delocalization with increasing disorder followed by Anderson localization upon further increasing disorder, was predicted~\cite{longhi2021inverse} and observed~\cite{li2022aharonov} in 1D systems.
It was also reported for potential disorder in \(d=3\) all band flat lattice~\cite{goda2006inverse, nishino2007flat}.
In that case the states exactly at the flat band energies remain metallic up to the Anderson localization transition. 
However there is a mobility edge in the spectrum and the states at the band edges are localized.
Therefore for a fixed energy different from the flatband energy, upon increasing the disorder, one first observes the localized states, then the delocalized states, until all the states become localized at sufficiently strong disorder.
We expect that a similar inverse Anderson transition might also occur in both the \(d=2\) SOCO case considered here and the \(d=3\) case studied in our previous work~\cite{cadez2021metalinsulator}.

\begin{acknowledgments}
    The authors acknowledge the financial support from the Institute for Basic Science (IBS) in the Republic of Korea through the project IBS-R024-D1.
\end{acknowledgments}

\appendix

\section{Clean ABF \texorpdfstring{\(\mh\)}{H} for \texorpdfstring{\(d = 1\) and \(d = 2\)}{1D and 2D}}
\label{app:fe_2d}

The ABF Hamiltonian \(\mh\) with \(\nu = 2\), and \(E_a = -1\), \(E_b = 1\), in the fully-entangled basis, is expressed as
\begin{align}
    \label{eq:fe}
    \mh = \sum_{\vect{r}} 
    \sum_{|\Delta \vect{r}|^2= 0}^{|\Delta \vect{r}|^2 = d}
    \hd^\dagger_{\vect{r}}H_{\Delta \vect{r}}\hd_{\vect{r} + \Delta \vect{r}}
\end{align}
where \(\hd_\vect{r} = (\hp_\vect{r} \,\, \hf_\vect{r})^T\). 
Note that, for the sake of convenience, here we use second quantization instead of the bra-ket notation.
The first sum is over all lattice sites \(\vect{r}\), and for each \(\vect{r}\), the second sum is for hoppings, where the hopping range is up to \(d\)-th nearest neighbors, 
and \(H_{\Delta \vect{r}}\) are the \(2 \times 2\) hopping matrices (See Fig.~\ref{fig:fig1} for intuitive understanding of the notation). 
The Hermiticity is ensured by the identity \(H_{-\Delta \vect{r}} = H^\dagger_{\Delta \vect{r}}\).
For \(d = 1\) we have
\begin{subequations}
    \label{eq:fe_1d_hop}
    \begin{align}
        \label{eq:fe_1d_hop1}
        & H_0 = \cos 2\theta 
        \begin{pmatrix}
            -\cos 2\theta & \phantom{-}\sin 2\theta \\
            \phantom{-}\sin 2\theta &  \phantom{-}\cos 2\theta
        \end{pmatrix} \\
        \label{eq:fe_1d_hop2}
        & H_1 = \frac{1}{2} \sin 2\theta 
        \begin{pmatrix}
            \sin 2\theta & -1 + \cos 2\theta \\
            1 + \cos 2\theta & -\sin 2\theta
        \end{pmatrix}.
    \end{align}
\end{subequations}
For \(d = 2\) we find
\begin{subequations}
    \begin{align}
    	\label{eq:fe_2d_1}
        & H_{0} = \cos^2 2\Th 
        \begin{pmatrix}
            -\cos 2\Th & \phantom{+}\sin 2\Th \\
            \phantom{+}\sin 2\Th    & -\cos 2\Th
        \end{pmatrix} \\
    	\label{eq:fe_2d_2}
        & H_{(1, 0)} =  \frac{1}{2}\sin 2\Th \cos 2\Th 
        \begin{pmatrix}
            \sin 2\Th & -1 + \cos 2\Th  \\
            1 + \cos 2\Th & -\sin 2\Th
        \end{pmatrix} \\
        \label{eq:fe_2d_3}
        & H_{(0, 1)} = \frac{1}{2}\sin^2 2\Th
        \begin{pmatrix}
            \cos 2\Th & -\sin 2\Th  \\
            -\sin 2\Th    & -\cos 2\Th
        \end{pmatrix} \\
        \label{eq:fe_2d_4}
        & H_{(1, 1)} = \frac{1}{4}\sin 2\Th (1+\cos 2\Th)
        \begin{pmatrix}
            \sin 2\Th  & -1 + \cos 2\Th \\
            1 + \cos 2\Th & - \sin 2\Th
        \end{pmatrix}\\
        \label{eq:fe_2d_5}
        & H_{(1, -1)} = \frac{1}{4} \sin 2\Th (1-\cos 2\Th)
        \begin{pmatrix}
            -\sin 2\Th & -1 - \cos 2\Th  \\
            1 - \cos 2\Th & \sin 2\Th
        \end{pmatrix}
    \end{align}
\end{subequations}
These hopping matrices correspond to nearest neighbor~(\ref{eq:fe_2d_2}-\ref{eq:fe_2d_3}), and second nearest neighbor~(\ref{eq:fe_2d_4}-\ref{eq:fe_2d_5}) hopping.
This is in accordance with the claim in Sec.~\ref{sec:model}, that the hoppings \(\mh\) extend up to \(d-\)th nearest neighbors.

\section{Finite size scaling analysis on \texorpdfstring{\(d = 2\)}{d=2} ABF with weak RMF disorder}
\label{app:fss_rmf}

The localization length is assumed to be a function of the logarithm of \(\bE\):
\begin{align}
    \xi = f(\ln |\bE|)
\end{align}
The \(\ttau\) is assumed to be a universal function of a single parameter, \(\chi = \ln L / \ln \xi\)
\begin{align}
    \ttau = F\left(\chi \right)
\end{align}
This ansatz is motivated by the finite size scaling analysis of hopping disordered bipartite lattices~\cite{markos2012disordered}.

With the above two assumptions, we fit our data using the Levenberg-Marquardt algorithm for nonlinear least squares fitting for the \(\ttau\) for \(\theta = 0.01 \pi\) in Fig.~\hyperlink{6a}{6a}. 
We restrict the energy range to \( 0.3 \leq \bE \leq 0.9\) because the above single parameter ansatz fails in the \(E = 0\) limit. This happens since the localization length diverges there, and hence \(F\) should be independent of \(L\), while \(\ttau\) is actually dependent on \(L\) at \(\bE = 0\).

\section{Finite size scaling analysis on \texorpdfstring{\(d = 2\)}{d=2} SOCO}
\label{app:fss_soco}

We follow the standard finite size scaling analysis as described in Ref.~\onlinecite{slevin2014critical}.
We assume the single parameter scaling \(\ttau(L, \theta) = F(L/\xi)\).
In the localized phase \(\xi\) is the localization length and in the extended phase \(\xi\) is a correlation length. 
We also assume that near the critical point \(\xi\) diverges as a power law, that is \(\xi \sim (\theta - \theta_c)^{-\nu}\) and that the scaling function \(F\) is analytical.
Defining \(F(L/\xi) = G(\phi_1)\), where \(\phi_1 = L^{1/\nu}u_1\) and \(|u_1| = \xi^{-1/\nu}\), we can Taylor expand both \(u_1\) and \(G\) as 
\begin{align}
    u_1 = \sum_{k=1}^{n} c_k (\theta - \theta_c)^k, \label{eq:C1} \\
    G(\phi_1) = \sum_{k=0}^{m} d_k \phi_1^k,
\end{align}
where \(u\) is a polynomial of degree \(n\) and \(G\) is a polynomial of degree \(m\), with the parameters to be determined being the critical exponent \(\nu\), the transition angle \(\theta_c\) and the coefficients of the Taylor expansion \(c_k, d_k\). The fitting is done by minimizing the \(\chi^2\) statistics using the Levenberg-Marquardt algorithm. The fitting results are summarized in TABLE~\ref{tb:rel}, where different orders of expansions \(n,m\) are used and we show the extracted value of the critical exponent \(\nu\), the minimized \(\chi^2\) value and the number of degrees of freedom \(N_{dof}\), which is the number of data points used in the fit minus the total number of parameters used. The results are consistent and give \(\nu = 2.55 \pm 0.03\), however the \(\chi^2 \sim 2 N_{dof}\) is relatively high, indicating that the data might be underfitted.

Thus we also consider a modified scaling, where \(G = G(\phi_1, \phi_2)\) is a function of \(\phi_1 = L^{1/\nu} u_1\), with \(u_1\) as given in eq.~\eqref{eq:C1} and \(\phi_2 = L^{y} u_2\), with \(u_2 = \sum_{k=0}^{i} \bar{c}_k (\theta - \theta_c)^k\) and \(y < 0\) being the irrelevant variable. The scaling function is Taylor expanded in both \(\phi_1\) and  \(\phi_2\) as
\begin{align}
G(\phi_1, \phi_2) = \sum_{k=0}^{m} \sum_{l=0}^{j} d_{k,l} \phi_1^k \phi_2^l,
\end{align}
and to avoid ambiguity we set \(d_{1,0} = d_{0,1} = 1\). The parameters to be fitted in this case are the critical exponent \(\nu\), the transition angle \(\theta_c\), the irrelevant variable \(y\) and the coefficients of the Taylor expansions \(c_k,\bar{c}_k, d_{k,l}\). As before we minimize the \(\chi^2\) statistics using the Levenberg-Marquardt algorithm and report the fitting results in TABLE~\ref{tb:irrel}.
Now the extracted critical exponent gives a larger value than above and has a larger error, the best result being \(\nu = 2.70 \pm 0.05\), however now the \(\chi^2 \sim  N_{dof}\) giving better fitting in the results obtained. The value of the critical exponent is in agreement with the known results in the literature~\cite{asada2002anderson, asada2004numerical}.


\begin{table}
    \begin{center}
        \begin{tabular}{|c|c|c|c|c|} 
             \hline
             $(n, m)$ & $ \theta_c $ & $\nu$ & $\chi^2$ & $N_{dof}$\\
             \hline\hline
             $(2,4)$ & $0.16100 \pm 7 \cdot 10^{-5}$ &$2.54\pm 0.03$ & 204 & 119 \\ 
             \hline
             $(2,5)$ & $0.16100 \pm 7 \cdot 10^{-5}$ &$2.55\pm 0.03$ & 203 & 118 \\ 
             \hline
             $(2,6)$ & $0.16094 \pm 7 \cdot 10^{-5}$ &$2.54\pm 0.03$ & 197 & 117 \\ 
             \hline
             $(3,6)$ & $0.1610 \pm 0.0001$ &$2.55\pm 0.04$ & 197 & 116 \\ 
             \hline
            $(4,6)$ & $0.1610 \pm 0.0001 $ 
 & $2.54\pm 0.07$ & 197 & 115 \\ 
            \hline
        \end{tabular}
    \end{center}
    \caption{
        Fitting results with only a relevant variable. 
    }
    \label{tb:rel}
\end{table}

\begin{table}
    \begin{center}
        \begin{tabular}{|c|c|c|c|c|} 
             \hline
             $(n, m, i, j)$ & $\theta_c$ & $\nu$ & $\chi^2$ & $N_{dof}$\\
             \hline\hline
             $(2, 4, 1, 2)$ & $0.1628 \pm 0.0006$ & $2.78 \pm 0.12$ & 105 & 114 \\ 
             \hline
             $(2, 4, 2, 2)$ &  $0.1621 \pm 0.0005$ & $2.70 \pm 0.05$ & 89 & 113 \\ 
             \hline
        \end{tabular}
    \end{center}
    \caption{
        Fitting results with irrelevant variable.
    }
    \label{tb:irrel}
\end{table}

\bibliography{general ,flatband, mbl, frustration}

\end{document}